# Variations on a classical theme: On the formal relationship between magnitudes per square arcsecond and luminance

Salvador Bará[1,*]

[11] Departamento de Física Aplicada, Universidade de Santiago de Compostela, 15782 Santiago de Compostela, Galicia, Spain.

**Abstract**

The formal link between magnitudes per square arcsecond and luminance is discussed in this paper. Directly related to the human visual system, luminance is defined in terms of the spectral radiance of the source, weighted by the CIE V(λ) luminous efficiency function, and scaled by the 683 lm/W luminous efficacy constant. In consequence, any exact and spectrum-independent relationship between luminance and magnitudes per square arcsecond requires that the latter be measured precisely in the CIE V(λ) band. The luminance value corresponding to $m_{VC}=0$ (zero-point of the CIE V(λ) magnitude scale) depends on the reference source chosen for the definition of the magnitude system. Using absolute AB magnitudes, the zero-point luminance of the CIE V(λ) photometric band is 10.96 x $10^4$ cd·m$^{-2}$.

## 1. Introduction

The quantitative evaluation of the night sky brightness is a matter of interest for a wide variety of studies, ranging from the naked-eye visibility of the stars and other celestial bodies to the disruptive effects of the anthropogenic skyglow on wildlife ecology and on potentially relevant aspects of human physiology. "Night sky brightness" is used here as a short-hand term for the spectral radiance of the night sky, angularly integrated within the field of view of the detector, and spectrally integrated over wavelengths, after being weighted by the spectral transmittance of the filter associated with the photometric band in which the measurements are taken.

The appropriate photometric band depends on the effects being studied: several sets of standard systems are currently used in astrophysics [1], like the well-known Johnson-Cousins UBVRI [2], whose zero-points can be defined either by the irradiance of a known physical source (e.g. the star Vega) or in terms of a prescribed spectral irradiance distribution, as the AB (absolute) magnitude system [3-4]. Environmental studies require the use of a wide set of specific action spectra and photometric bands [5-12], to account for the variety of physiological and behavioural processes of the different species sharing an artificially illuminated area. Human visual perception, in turn, is described using the standard CIE V(λ) luminous efficiency function [13-14], which informs us of the relative sensitivity of the visual system to the different wavelengths of the optical spectrum. The standard CIE V(λ) function is defined as the observer-averaged inverse of the relative radiant power required to elicit a perception of equal brightness, in foveal vision, of two adjacent (or secuentially presented, in flicker mode) uniform light fields of different wavelength, one of which is taken as the reference.

Besides in the linear radiance scale of units Wm$^{-2}$sr$^{-1}$ (or luminance, in cd·m$^{-2}$, in case the measurement band coincides strictly with the CIE V(λ) one), the night sky brightness can also be expressed in the logarithmic scale of magnitudes per square arcsecond, traditionally used in astrophysics. Many practical devices currently used in light pollution studies provide the output in these units. The all-sky system ASTMON [15], for instance, reports angularly resolved measurements of the brightness across the celestial hemisphere above the observer, in magnitudes per square arcsecond in the Johnson-Cousins B, V, and R bands. The widely-used low-cost radiometers SQM (Unihedron, Canada) [4,16-19] and TESS [20] use device-specific photometric bands. Other popular detectors, like the DSLR cameras, provide radiance readings in the RGB color space, from which the Johnson V magnitudes can be approximately estimated. Not infrequently, a loose identification is made for practical purposes among the Johnson V, the SQM, and the CIE V(λ) bands, in spite of the potentially strong differences between the brightnesses of the same source reported in each one of them, that have been comprehensively analyzed and discussed by Cinzano [16], and especially by Sánchez de Miguel et al. [4]. Several approximate expressions can be found in the literature for transforming magnitudes per square arcsecond measured in the Johnson V, or in the specific SQM band, into luminance in cdm$^{-2}$, with zero-point factors in the interval $9.0 - 10.9 \times 10^4$ cdm$^{-2}$. Some of them are analyzed in Subsection 3.2 below.

* Salvador Bará
*E-mail address:* salva.bara@usc.es

The purpose of this communication is to describe the formal relationship between magnitudes per square arcsecond and luminance, a relationship that can only be established with exact and general (spectrum-independent) validity if the magnitudes are measured and reported in the CIE V(λ) photometric band.

## 2. Luminance and AB magnitudes per square arcsecond in the CIE V(λ) band

Astronomical magnitudes are a relative negative logarithmic scale commonly used in astrophysics for reporting the irradiance [Wm$^{-2}$] in any predefined photometric band. Denoting by $T(\lambda)$ the spectral weighting function (filter) characteristic of that band, the magnitude $m$ of a source is defined as:

$$m = -2.5\log_{10}\left[\frac{\int_{\lambda=0}^{\infty} T(\lambda)E(\lambda)\,\mathrm{d}\lambda}{\int_{\lambda=0}^{\infty} T(\lambda)E_0(\lambda)\,\mathrm{d}\lambda}\right], \tag{1}$$

where $E(\lambda)$ [Wm$^{-2}$nm$^{-1}$] is the spectral irradiance produced by the source at the entrance plane of the filter, and $E_0(\lambda)$ is a spectral irradiance distribution chosen as the reference for determining the origin (zero-point) of the $m$ magnitude scale. According to Eq.(1) the overall filter-weighted irradiance in the $T$ band, $E_T = \int_{\lambda=0}^{\infty} T(\lambda)E(\lambda)\,\mathrm{d}\lambda$, can be expressed in terms of the reference one, $E_{0,T} = \int_{\lambda=0}^{\infty} T(\lambda)E_0(\lambda)\,\mathrm{d}\lambda$, using the well-known formula

$$E_T = E_{0,T} \times 10^{-0.4m}. \tag{2}$$

Magnitudes per square arcsecond, in turn, are a non-SI scale for reporting the filter-weighted radiance [Wm$^{-2}$sr$^{-1}$] incident on the detector. Let us remind that, by definition, the spectral irradiance $E(\lambda)$ and the spectral radiance $L(\lambda;\boldsymbol{\alpha})$ [Wm$^{-2}$sr$^{-1}$nm$^{-1}$] are related through the equation [21]

$$\mathrm{d}E(\lambda) = L(\lambda;\boldsymbol{\alpha})\cos\theta\,\mathrm{d}\omega, \tag{3}$$

where $\mathrm{d}E(\lambda)$ is the elementary irradiance produced by a beam of radiance $L(\lambda;\boldsymbol{\alpha})$ incident on the detector from an infinitesimal cone of directions $\mathrm{d}\omega$ (elementary solid angle) around the direction $\boldsymbol{\alpha}$, which forms an angle $\theta$ with the normal to the detector surface. If directions in space are described using spherical coordinates, then $\boldsymbol{\alpha} = (\phi,\theta)$, being $\phi$ the azimuth and $\theta$ the polar angle, and $\mathrm{d}\omega = \sin\theta\,\mathrm{d}\theta\,\mathrm{d}\phi$.

If the angular size of the source, $\Delta\omega$, is small and the detector is pointing directly towards it ($\theta \cong 0$), then $L(\lambda;\boldsymbol{\alpha}) \cong L(\lambda)$ and $\cos\theta \cong 1$, and hence $L(\lambda) \cong \Delta E(\lambda)/\Delta\omega$. This finite angular approximation can be safely applied to compute the spectral radiance of one square arcsecond patch of the sky, since 1 arcsec$^2$=$2.3504 \times 10^{-11}$ sr.

The statement that a uniform extended source has a brightness of $m_T$ magnitudes per square arcsecond in the $T$ band is equivalent to the statement that the filter-weighted irradiance $\Delta E$ produced at the detector plane by one square arcsecond patch of that source is equal to the filter-weighted irradiance $E_T$ that would be produced at the same plane by a point source of magnitude $m = m_T$, acording to Eq.(2). Hence the radiance $L_T$ corresponding to a brightness of $m_T$ magnitudes per square arcsecond in the $T$ band is

$$L_T = \left(\frac{E_{0'T}}{\Delta\omega_0}\right) \times 10^{-0.4 m_T} \equiv L_{0,T} \times 10^{-0.4 m_T} \qquad [\mathrm{Wm^{-2}sr^{-1}}] \qquad (4)$$

with $\Delta\omega_0 = 2.3504 \times 10^{-11}$ sr.

The above expressions can be particularized for the the CIE V(λ) measurement band, by substituting $V(\lambda)$ for $T(\lambda)$ and the subindex "*VC*" for "*T*" where appropriate. In this case (and only in this case) the irradiance $E_{VC}$ in Eq.(2) can be equivalently expressed as an illuminance $E'_{VC}$, in SI visual units lx=lm/m$^2$, and the radiance $L_{VC}$ in Eq.(4) can be expressed as a luminance $L'_{VC}$, in units cdm$^{-2}$, or their equivalent, lx·sr$^{-1}$. This is achieved by multiplying both expressions by the standard luminous efficacy scaling factor 683 lm/W. The luminance $L'_{VC}$ is then given by

$$L'_{VC} = 683 \left[lmW^{-1}\right] \times \left(\frac{E_{0'VC}}{\Delta\omega_0}\right) \times 10^{-0.4 m_{VC}} \equiv L'_{0,VC} \times 10^{-0.4 m_{VC}} \qquad [\mathrm{cdm^{-2}}] \qquad (5)$$

with a zero-point, in luminous units,

$$L'_{0,VC} = \frac{683 \left[lmW^{-1}\right]}{\Delta\omega_0} \times \int_{\lambda=0}^{\infty} V(\lambda) E_0(\lambda) \mathrm{d}\lambda \qquad [\mathrm{cdm^{-2}}] \qquad (6)$$

The precise value of $L'_{0,VC}$ is contingent upon the choice of the spectral irradiance distribution $E_0(\lambda)$ that will be used as a reference to set the zero-point of the magnitude scale. A possible choice is the spectral irradiance of the A0V type star Vega ($\alpha$ Lyr), of effective temperature $T_e$=9550 K and Johnson V-band magnitude +0.03 [22], with null Johnson color indices U−B=B−V=0. An alternative to this choice, not tied to any particular reference star, is the AB (absolute) magnitude scaling [3-4], whereby the reference spectral irradiance, per unit frequency interval, is set to a constant value of $E_0(\nu)$=3631 Jansky (Jy) throughout the whole spectral domain (1 Jy = $10^{-26}$ Wm$^{-2}$Hz$^{-1}$). Taking into account that $E_0(\lambda)\mathrm{d}\lambda = E_0(\nu)\mathrm{d}\nu$, where $\mathrm{d}\nu$ is the frequency spectral interval (in Hz) corresponding to the wavelength interval $\mathrm{d}\lambda$ (in m), and that $\nu = c/\lambda$, so that $\mathrm{d}\nu = \left(-c/\lambda^2\right)\mathrm{d}\lambda$, the reference AB spectral irradiance per unit wavelength interval is given by

$$E_0(\lambda) = 3631 \left[Jy\right] \times \left(c/\lambda^2\right), \qquad [\mathrm{Wm^{-2}m^{-1}}] \qquad (7)$$

with $c$= 299,792,458 ms$^{-1}$ and $\lambda$ in m. Note that the right-hand side of Eq.(7) shall be multiplied by $10^{-9}$ in case $E_0(\lambda)$ is expressed in Wm$^{-2}$nm$^{-1}$. The minus sign in the expression relating $\mathrm{d}\nu$ and $\mathrm{d}\lambda$ cancels out, since the change from the frequency to the wavelength domain involves a change of the integration limits in Eqs.(1) and (6) that introduces an additional sign reversal. Eq. (6) becomes then:

$$L'_{0,VC} = 683 \left[lmW^{-1}\right] \times 3631 \left[Jy\right] \times \frac{c}{\Delta\omega_0} \times \int_{\lambda=0}^{\infty} \frac{V(\lambda)}{\lambda^2} \mathrm{d}\lambda \,. \qquad (8)$$

Computing numerically the integral in Eq.(8) we get the zero-point:

$$L'_{0,VC} = 10.96 \times 10^4 \,, \qquad [\mathrm{cd \cdot m^{-2}}] \qquad (9)$$

so that, finally,

$$L'_{VC} = 10.96 \times 10^4 \times 10^{-0.4 m_{VC}} \,. \qquad [\mathrm{cd \cdot m^{-2}}] \qquad (10)$$

Other constants of interest for the AB magnitude scale in the CIE V(λ) band are listed in Table 1. They were computed using the CIE 2º photopic luminosity curve (1924) V(λ) numeric values from [23], linearly interpolated with 1 nm resolution. The integrations were performed in the interval 340-780 nm.

Table 1. Constants of the AB magnitude scale in the CIE V(λ) band

| Constant | Symbol | Value | Units |
|---|---|---|---|
| Zero-point irradiance | $E_{O,VC}$ | 3.77 x $10^{-9}$ | Wm$^{-2}$ |
| Zero-point radiance | $E'_{O,VC}$ | 160.42 | Wm$^{-2}$sr$^{-1}$ |
| Zero-point iluminance | $L_{O,VC}$ | 2.58 x $10^{-6}$ | lx |
| Zero-point luminance | $L'_{O,VC}$ | 10.96 x $10^4$ | cd·m$^{-2}$ |

## 3. Discussion

### *3.1. Additional remarks*

The previous section provides a simple deduction of the relationship between luminance, in cdm$^{-2}$, and magnitudes per square arcsecond. This relationship is accurate and has general validity if -and only if- the magnitudes are measured in the CIE V(λ) band that is at the root of the luminance definition. The choice of the zero-point of the magnitude scale is however arbitrary, and several options can be chosen to specify the reference spectral irradiance. In this work we have used the AB magnitude system, based on the specification of a constant spectral irradiance of 3631 Jy throughout the whole spectral range. This choice presents the advantages of being absolute and not tied to the measurement of the spectral irradiance of any reference source, as, e.g. the star Vega ($\alpha$ Lyr) or the Sun.

Note that the use of the CIE V(λ) band is a requirement in order to be consistent with the accepted definition of luminance. Luminance is a standard photometric magnitude closely related to the human perception of brightness in foveal vision. The V(λ) function for a 2º field of view basically expresses the combined spectral response of the foveal L and M cones, with a small contribution from the (mainly extrafoveal) S cone photoreceptors. Several modifications of the classical V(λ) function have been developped by Judd and Vos [24], and Sharpe and Stockmann [25], motivated in part for the need of correcting the too-low value of the CIE V(λ) function at the short wavelength range of the visible spectrum (datasets available in [23]). The V(λ) function for extended fields of view (10º) [23] may also be utilised to overcome this drawback. A recent model proposed by Rea et al. [26] introduces a linear combination of the V(λ) and the S(λ) cone sensitivity for describing the wide-field perception of brightness, as opossed to the foveal one. Besides, the spectral response of the human eye also depends on its state of adaptation to the prevailing luminance level. For dark-adapted eyes the CIE 1951 V'(λ) scotopic sensitivity function [27] should be used instead of the photopic V(λ) one. For intermediate luminance adaptation levels, several mesopic sensitivity functions have been proposed [28-29]. None of these functions have been used here, to be consistent with the classical luminance definition. Further studies, however, may suggest the use of any of them as the band of choice for better characterizing the human perception of brightness of the night sky.

Note also that the definition of the magnitude system can be made using spectral energy densities, as in Eq.(1) above, or spectral photon number densities, as in Eq.(9) of [30]. Both definitions, although not strictly equivalent, only differ by minor amounts for most practical situations.

As deduced from the previous paragraphs, and has been thoroughly analyzed by Sánchez de Miguel et al. [4], it is not possible to define an exact, general (spectrum-independent), and unambiguous transformation between luminance and magnitudes per square arcsecond if the magnitudes are measured in bands different from the CIE V(λ). Any particular conversion attempt shall specify the offset between the measured magnitudes and the V(λ) ones. This offset turns out to be spectrum-dependent, so that only by knowing the spectral distribution of the source is it possible to compute it with accuracy. That being said, for light pollution studies under well-characterized circumstances one may get reasonably good estimations of the conversion factor, since the spectral distributions of the main types of artificial lighting sources are relatively well-known and belong to a restricted set of technologies.

### *3.2. Some notes on previously published formulae*

Several formulae can be found in the literature relating the luminance of an extended source to its brightness in magnitudes per square arcsecond in the Johnson V band, that will henceforth be denoted by $m_{VJ}$ in order to avoid confusion with the CIE V(λ) $m_{VC}$. As it is evident from the above section and from previously published works [3-4], these conversion formulae must be considered only approximate, since sources with the same $m_{VJ}$ will generally have different $m_{VC}$, and consequently will give rise to different luminances, depending on their particular spectral composition.

An expression of this kind frequently used in light pollution studies (see, e.g. [31-35]) is

$$L'_{VC} \cong 10.8 \times 10^{4} \times 10^{-0.4 m_{VJ}} \equiv \hat{L}'_{0,VC} \times 10^{-0.4 m_{VJ}} \qquad [\mathrm{cd \cdot m^{-2}}]. \qquad (11)$$

There has been some discussion about the precise origin and range of validity of this formula. Several recent [36] and classical [37-38] works quote the following expression by Garstang [39] to transform the so-called "visual magnitude" of the sky, *V*, given in Johnson magnitudes per square second ( $m_{VJ}$ , in our notation), to brightness (*b*) expressed in nanolamberts (1 nL = $10^{-5}/\pi$ cd·m$^{-2}$):

$$b = 34.08\ \exp\left(20.7233\ -\ 0.92104\ V\right) \qquad [\text{nL}]. \qquad (12)$$

Transforming the base-*e* exponentials into base-10, and expressing the nL in cd·m$^{-2}$, Eq.(12) can be rewritten in the form of Eq.(11) with a zero-point luminance $\hat{L}'_{0,VC} = 10.85\times10^{4}$ cd·m$^{-2}$. In ref [39], Eq. (12) is said to be based on the conversion given by Allen in p. 26 of [40], that establishes the value of the luminance of "one $m_v$=0 star per square degree outside the atmosphere" as $2.63\times10^{-6}$ lambert (L), with no further indication about its derivation. Transforming the lamberts to cd·m$^{-2}$ and the square degrees to square arcseconds, this value gives an Allen´s zero-point luminance $\hat{L}'_{0,VC} = 10.89\times10^{4}$ cd·m$^{-2}$, in agreement with the value $10.9\times10^{4}$ cd·m$^{-2}$ quoted by Slychter [41].

In a different section of Allen's book, the "illuminance of one $m_v$=0 star outside the atmosphere" is quoted as $\hat{E}'_{0,VC} = 2.54\times10^{-6}$ lx ([40] p. 197). For small angular sources this illuminance corresponds to a zero-point luminance $\hat{L}'_{0,VC} = 10.81\times10^{4}$ cd·m$^{-2}$. No detailed explanations are given about the precise origin of this value of the extra-atmospheric zero-point illuminance for a $m_{VJ}$ =0 star, but, as a matter of fact, it is consistent with the expected illuminance produced by a $m_{VJ}$ =0 star with a Vega-like spectral distribution, if the original Johnson V zero-point irradiance is rescaled such that the $m_{VJ}$ magnitude of Vega is precisely +0.00. The original absolute calibration of the Johnson V band [42] was made by setting the Sun magnitude to $m_{VJ}$ =–26.74 (+/– 0.05), and computing the absolute V band weighted irradiance of the Sun using Allen's solar spectral irradiance distribution provided in p. 172 of [40], which is an adaptation of the Labs and Neckel classical Sun spectral dataset [43]. This original calibration gives $E_{0,VJ} = 3.30\times10^{-9}$ Wm$^{-2}$ and $L_{0,VJ} = 140.6$ Wm$^{-2}$sr$^{-1}$ for the radiometric zero-point constants of the Johnson V band. In this system the Vega magnitude is $m_{VJ}$ = +0.03. If the above constants are rescaled such that $m_{VJ}$ (Vega)=0.00 (this amounts to multiplying their values by $10^{-0.4\times0.03}$ ), the new zero-point constants become $E_{0,VJ} = 3.21\times10^{-9}$ Wm$^{-2}$ and $L_{0,VJ} = 136.8$ Wm$^{-2}$sr$^{-1}$. A Vega-like source characterized by a 9550 K blackbody spectrum [22] and this value of the $E_{0,VJ}$ irradiance would give rise to a CIE V(λ) illuminance $\hat{E}'_{0,VJ} = 2.55\times10^{-6}$ lx and an associated luminance $\hat{L}'_{0,VJ} = 10.84\times10^{4}$ cd·m$^{-2}$, in overall agreement with Allen's value and with the one in Eq.(11). This equation can then be applied for transforming magnitudes to luminance, provided that the source has a blackbody spectral radiance distribution with effective temperature 9550 K, and the magnitudes are measured in the Johnson V band with a zero-point defined by $m_{VJ}$ (Vega)=0.00. However, the zero-point luminance shall be modified if the source is a blackbody of different temperature, reaching a value of $\hat{L}'_{0,VJ} = 12.25\times10^{4}$ cd·m$^{-2}$ for 2500 K sources. For other types of spectra the corresponding corrections can be deduced from the results of Sanchez de Miguel et al [4].

The numerical values in this section have been obtained using the Johnson V weighting function given by Bessel in Table 2 of [44], interpolated to 1 nm resolution. Integrations were carrried out in the interval 340-780 nm.

## 4. Conclusions and recommendations

The formal relationship between luminance and magnitudes per square arcsecond in the CIE V(λ) band is described in this work, specifying the zero-point radiometric and photometric constants of this band in the AB magnitude scale. This relationship provides an accurate and general (spectrum-independent) link between the linear SI visual luminous scale and the magnitude one.

This conversion holds provided that the magnitudes are measured precisely in the CIE V(λ) band. Magnitudes per square arcsecond measured in other photometric bands, as is widely known, cannot be unambiguously converted into luminances if the spectral radiance distribution of the source is unknown. Notwithstanding that, for many applications in light pollution research the visual brightness of the night sky must be measured and reported, even if only approximately, using the detectors at hand. Whereas specific low-cost high-sensitivity radiometers operating in the CIE $V(\lambda)$ band will not be generally available, the use of magnitudes measured in other photometric bands (Johnson V or device-specific ones) may allow, using approximate conversion formulae, to get some insight into the visual brightness of the night sky, with expected deviations below a few tens percent.

For quantitative light pollution studies, however, absolute radiometric values should also be reported when practicable. Many detector devices provide raw data that are directly convertible to band-weighted irradiances or radiances after a suitable absolute calibration: these radiometric values have a direct physical significance, an unambiguous meaning, and should ideally be disclosed in research publications. Equivalently, if magnitudes per square arcsecond are the units of choice, the radiometric constants of the zero-point of the magnitude scale, in SI units $Wm^{-2}$ or $Wm^{-2}sr^{-1}$, should be explicitly given.

**Acknowledgements**

This work was partially developed within the framework of the Spanish Network for Light Pollution Studies (AYA2015-71542-REDT).